\documentclass{moriond}

\usepackage{amsmath}
\def\be{\begin{equation}}
\def\ee{\end{equation}}
\def\bea{\begin{eqnarray}}
\def\eea{\end{eqnarray}}

\newcommand{\CHlt}{C_{Hl}^{(3)}}
\newcommand{\CHqt}{C_{Hq}^{(3)}}
\newcommand{\CHls}{C_{Hl}^{(1)}}
\newcommand{\CHqs}{C_{Hq}^{(1)}}

\begin{document}
\vspace*{4cm}
\title{EWPD in the SMEFT and the $\mathcal{O}(y_t^2,\lambda)$ one loop $Z$ decay width}

\author{M.Trott}

\address{Niels Bohr International Academy \& Discovery Center,\\
Niels Bohr Institute, University of Copenhagen,\\
Blegdamsvej 17, DK-2100, Copenhagen, Denmark}

\maketitle\abstracts{The interpretation of electroweak precision data in the Standard Model Effective Field Theory is discussed. One loop corrections
$\propto y_t^2, \lambda$ to the partial $Z$ decay widths and ratios of partial widths
in this theory are discussed. A reparameterization invariance
and the non-minimal character of matching onto this theory is reviewed.}

\section{Introduction}

Electroweak precision data (EWPD) provides information on the interactions of the known Standard Model (SM) particles
around the Electroweak (EW) scale. These measurements supply a consistency test
for the SM or any model that seeks to extend or supplant the SM.
It is crucial to combine this information in a consistent manner with the measurements of the
properties of the Higgs-like ($J^P = 0^+$) scalar discovered at the Large Hadron Collider (LHC)
to determine the global constraint picture on the SM, or physics beyond the SM.

When non-SM interactions and states are associated with scales
parametrically separated from the EW scale ($\Lambda \gg v$) one can Taylor expand
the effects of $\Lambda$ on lower energy measurements. For EWPD measured
on the $W,Z$ poles, the momentum scales are effectively limited to
$p^2 \simeq m_W^2,m_Z^2$.
Non-analytic structure of the full correlation functions describing these observables is only generated by the SM states;
the new states extending the SM cannot go on-shell as $\Lambda \gg v$. As a result, one can expand in the ratio $v^2/\Lambda^2$
the effect of this unknown physics into a series of analytic
and local operators, generating the Standard Model Effective Field Theory (SMEFT)
\bea
\mathcal{L}_{SMEFT} = \mathcal{L}_{SM} + \mathcal{L}^{(5)} + \mathcal{L}^{(6)} + \mathcal{L}^{(7)} + ...,
\quad \quad \mathcal{L}^{(d)}= \sum_{i = 1}^{n_d} \frac{C_i^{(d)}}{\Lambda^{d-4}} Q_i^{(d)} \hspace{0.25cm} \text{ for $d > 4$.}
\eea
The operators  $Q_i^{(d)}$ are suppressed by $d-4$ powers of the cutoff scale $\Lambda$,
where the $C_i^{(d)}$ are the Wilson coefficients.
The number of non redundant operators in $\mathcal{L}^{(5)}$, $\mathcal{L}^{(6)}$, $\mathcal{L}^{(7)}$ and $\mathcal{L}^{(8)}$ is
known \cite{Buchmuller:1985jz,Grzadkowski:2010es,Weinberg:1979sa,Wilczek:1979hc,Abbott:1980zj,Lehman:2014jma,Lehman:2015coa,Henning:2015alf}.
The SMEFT takes advantage of this drastic simplification in how the effects of physics beyond the SM can
appear, due to a separation of scales.
The SMEFT separates the description of the processes under study
into the long distance (infrared or IR) propagating states and their interactions, captured by
the operator expansion, and the ultraviolet (or UV) dependent Wilson coefficients. A model
independent analysis treats these Wilson coefficients as free parameters to be fit from the data, subject
to the constraint that the operator expansion is convergent (i.e. $C_i \, v^2/\Lambda^2 < 1$) to retain a predictive theory.
By making the IR assumption to include a Higgs doublet in the EFT, the SMEFT results from this Taylor expansion.

A large number of studies on EWPD, Higgs data,
and the combination thereof, have been done in the SMEFT.
Many of these studies are correct despite the conflicted literature. The different conclusions found are due
to different analysis choices, the treatment (or neglect) of theoretial errors
in the various works, and different UV assumptions.
The less model independent conclusions, that neglect essential theoretical errors,
are more constrained.
The results presented here aim to develop a more comprehensive and model independent
understanding of the constraints of EWPD measurements projected onto the SMEFT.

In this proceeding, the general constraint picture in EWPD is reviewed in Section \ref{subsec:generalities}.
The utility of the SMEFT to examine and quantify measurement bias when interpreting the data outside of
the SM is discussed in Section \ref{subsec:bias}. The effect of loop corrections $\propto y_t^2, \lambda$ is discussed in Section \ref{subsec:loops}.
A subtle reparameterization invariance that explains the
highly correlated Wilson coefficient space in the SMEFT is discussed in Section \ref{subsec:reparam}.
Finally, some comments on the non-minimal character of
the SMEFT and universal theories are made in Section \ref{subsec:nonminimal}.

\section{General SMEFT constraint picture}\label{subsec:generalities}
Most analyses of EWPD are still performed
using the $\rm S,T$ formalism, which parameterizes a few common corrections to the two point functions of the gauge bosons
($\Pi_{WW,ZZ,\gamma Z}$) as\cite{Agashe:2014kda}
\bea\label{Sdefn}
\frac{\hat{\alpha}(m_Z)}{4 \, \hat{s}_Z^2 \, \hat{c}_Z^2} \, S &\equiv&
\frac{\Pi_{ZZ}^{new}(m_Z^2) - \Pi_{ZZ}^{new}(0)}{m_Z^2}
 - \frac{\hat{c}_Z^2-\hat{s}_Z^2}{\hat{c}_Z \, \hat{s}_Z} \frac{\Pi^{new}_{Z \, \gamma}(m_Z^2)}{m_Z^2} -  \frac{\Pi^{new}_{\gamma \, \gamma}(m_Z^2)}{m_Z^2}, \\
\hat{\alpha} T &\equiv& \frac{\Pi_{WW}^{new}(0)}{m_W^2}-\frac{\Pi_{ZZ}^{new}(0)}{m_Z^2}.
\eea
One calculates $\Pi_{WW,ZZ,\gamma Z}$ in a model and uses global fit results on $\rm S,T$ to constrain the model. This can only be done if
the conditions on the global $\rm S,T$ EWPD fit are satisfied; that vertex corrections due to physics beyond the SM are neglected - giving the ``oblique" qualifier \cite{Peskin:1991sw}.
The SM Higgs couples in a dominant fashion to $\Pi_{WW,ZZ}$
when generating the mass of the $W,Z$ bosons, and has small couplings to the light fermions, satisfying the oblique assumptions.
LHC results indicate the
$W,Z$ bosons obtain their mass in a manner that is associated with the Higgs-like scalar.
Corrections to $\Pi_{WW,ZZ}$ can be included for the SM,
or more generally\cite{Barbieri:2007bh} due to this scalar. Once this is done,
there is no strong theoretical support to maintain an oblique assumption using EWPD to constrain new physics
scenarios. Transitioning
away from this assumption to a SMEFT analysis permits the determination of higher order corrections when interpreting EWPD,
see Section \ref{subsec:loops}. Finally, the essential problem overcome by adopting a consistent SMEFT analysis
is that the oblique assumption is not field redefinition invariant.
The SM equation of motion (EOM) for the gauge fields are
\begin{eqnarray}\label{EOMsm}
\left[D^\alpha , W_{\alpha \beta} \right]^I &=& g_2   \frac 12 \overline q \, \tau^I \gamma_\beta  q + \frac12 \overline l \, \tau^I \gamma_\beta  l
+ \frac12 H^\dagger \, i\overleftrightarrow D_\beta^I H\, \\
D^\alpha B_{\alpha \beta} &=& g_1  \sum_{\psi}
\overline \psi \, y_i \gamma_\beta  \psi +
\frac12 H^\dagger \, i\overleftrightarrow{D}_\beta H.
\end{eqnarray}
Where $H^\dagger \, i\overleftrightarrow D_\beta H = i H^\dagger (D_\beta H) - i (D_\beta H)^\dagger H$,
$\psi=\{u,d,q,e,l\}$, $\tau^I$ is the Pauli matrix,
and $H^\dagger \, i\overleftrightarrow D_\beta^I H = i H^\dagger \tau^I (D_\beta H) - i (D_\beta H)^\dagger\tau^I H$.
A change of variables in the path integral
can be used to map the effects of physics beyond the SM represented in the SMEFT from $\Pi_{WW,ZZ,\gamma Z}$, to vertex corrections,
as the different terms in Eqns.4,5 transform the same way under $\rm SU(3) \times SU_L(2) \times U_Y(1)$. In the $S,T$ approach, some
of these corrections are retained, and others are neglected by assumption.
Deviations characterised by higher dimensional operators in the SMEFT
are described in a operator basis chosen by using the EOM (including Eqns.4,5), to reduce from
an over-complete Lagrangian to a reduced non-redundant basis.
Attempts to translate the oblique condition into a requirement to use
a particular operator basis by using these EOM relations, are afflicted with terminal internal inconsistencies,
and limited to describing UV scenarios sometimes known as ``universal theories''\cite{Barbieri:2004qk}. See Section \ref{subsec:nonminimal} for more discussion.

\begin{figure}
 \centering
 \includegraphics[width=\textwidth]{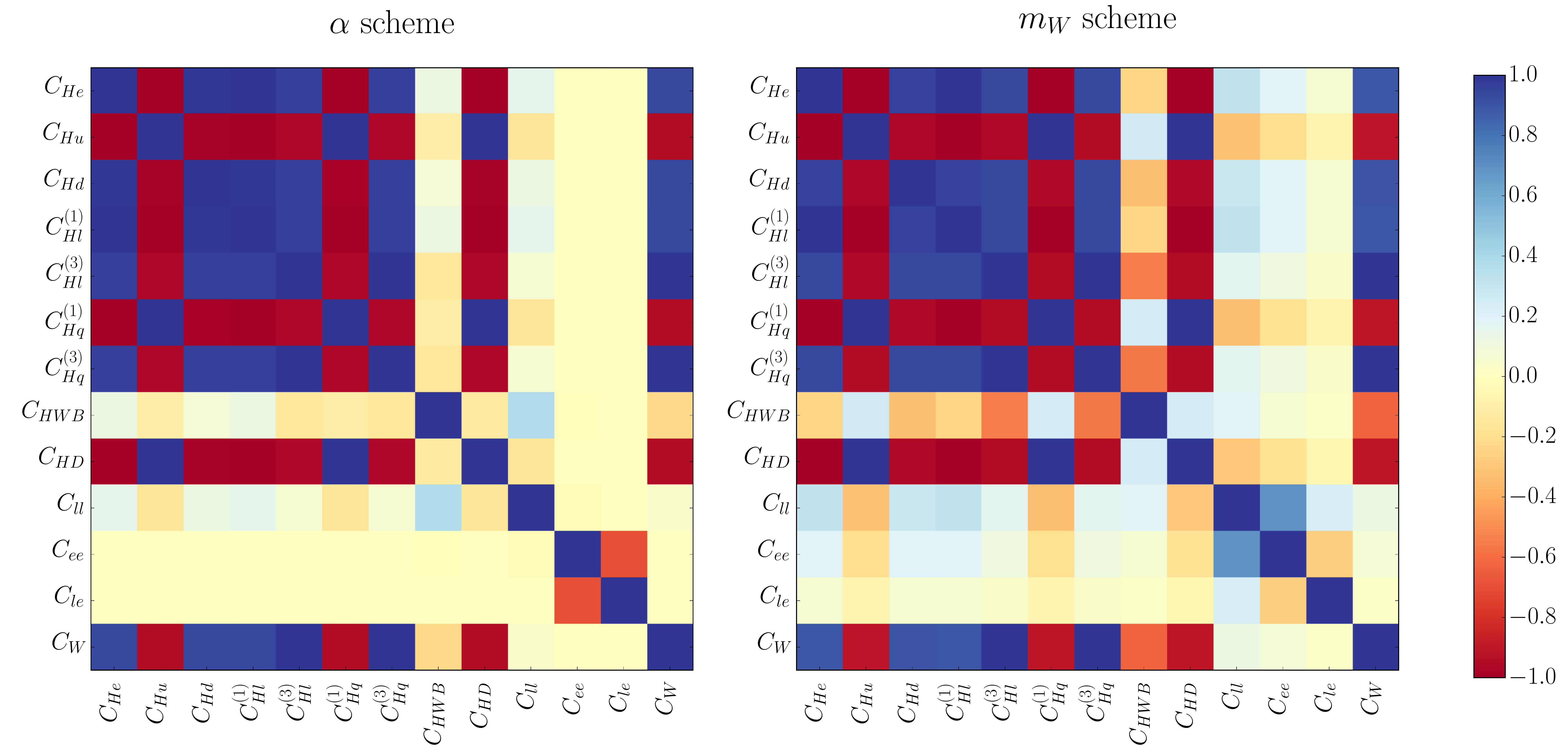}
 \caption{\label{Fig:correlation_matrices}Color map of the correlation matrix among the Wilson coefficients,
 obtained assuming zero SMEFT error, for the $\{\hat{\alpha}, \hat{m}_Z,\hat{G}_F\}$ input scheme (left)
 and for the $\{\hat{m}_W, \hat{m}_Z,\hat{G}_F\}$ input scheme (right).}
\end{figure}
Dropping these assumptions leads to the SMEFT analysis of EWPD. First a
model is mapped to the SMEFT Wilson coefficients in a tree or loop level matching calculation.
Then model independent global fit results are used to constrain the Wilson coefficients.
Works to analyse EWPD with a focus on EFT methods in a
proto-SMEFT setting appeared long ago \cite{Grinstein:1991cd,Han:2004az}.
The analysis of Han \& Skiba\cite{Han:2004az} identified
unconstrained directions in the EWPD set, and maintained a sober judgement of the degree of constraint
on the highly correlated Wilson coefficient space. Recent analyses still find
that the EWPD Wilson coefficient space is highly correlated.\cite{Berthier:2015oma,Berthier:2015gja,Falkowski:2014tna,Berthier:2016tkq,Brivio:2017bnu} It is required
to include data from $\bar{\psi} \psi \rightarrow \bar{\psi} \psi \bar{\psi} \psi$ scattering to lift the flat directions
present in the Wilson coefficient space\cite{Han:2004az}, for example when using the warsaw operator basis \cite{Grzadkowski:2010es}
as can be understood to follow from a reparameterization invariance, see Section \ref{subsec:reparam}.
The Wilson coefficients of the warsaw basis operators are labeled as $C_i$ and we refer the reader to this reference
for the explicit operator definitions.
In determining the constraints on the Wilson coefficients of the SMEFT,
one chooses an input parameter set, and predicts
EWPD observables.
In the $Z,W$ pole results\cite{Berthier:2015oma,Berthier:2015gja} the mapping is
\bea
\{\hat{m}_Z,\hat{m}_h,\hat{m}_t,\hat{G}_F,\hat{\alpha}_{ew},\hat{\alpha}_s,\Delta \hat{\alpha} \}
\rightarrow \{m_W,\sigma_h^0,\Gamma_Z,R_\ell^0,R_b^0,R_c^0,A_{FB}^\ell,A_{FB}^c,A_{FB}^b\},
\eea
through $\mathcal{L}_{SMEFT}$. Here the hat suprescript indicates an input parameter.
A SMEFT fit procedure is as follows.  A set of observables
is denoted $O_i$, $\bar{O}^{LO}_i$, $\hat{O}_i$ for the SM prediction, SMEFT prediction to first order in the $C^{(6)}$,
and the experimental value respectively.
Assuming the measured value $\hat{O}_i$ to be a gaussian variable centred about $\bar{O}_i$,
the likelihood function ($L(C)$) and $\chi^2$ are
\bea
L(C) = \frac{1}{\sqrt{(2 \pi)^n |V|}} \text{exp} \left(-\frac{1}{2} \left( \hat{O} - \bar{O}^{LO}\right)^T V^{-1} \left( \hat{O} - \bar{O}^{LO}\right)\right),
\quad \quad \chi^2 = - 2 \text{Log}[L(C)],
\eea
where $V_{ij} = \Delta^{exp}_i \rho^{exp}_{ij} \Delta^{exp}_j + \Delta^{th}_i \rho^{th}_{ij} \Delta^{th}_j$
is the covariance matrix with determinant $|V|$.
$\rho^{exp}$/$\rho^{th}$ are the experimental/theoretical correlation matrices
and $\Delta^{exp}$/$\Delta^{th}$ the experimental/theoretical error of the observable $O_i$.
This approach is necessarily an approximation, with neglected
effects introducing a theoretical error.\cite{Berthier:2015oma,Berthier:2015gja,David:2015waa}
The theoretical error $\Delta_i^{th}$ for an observable $O_i$ is defined as
$\Delta^{th}_i = \sqrt{\Delta_{i,SM}^2 + \left(\Delta_{i,SMEFT} \times O_i\right)^2}$,
where $\Delta_{i,SM}$, $\Delta_{i,SMEFT}$ correspond to the absolute SM theoretical, and the multiplicative SMEFT theory error.
The $\chi^2$ is
\bea
\chi^2_{C_i^6} = \chi^2_{C_i^6, min} + \left(C_i^6 - C_{i,min}^6 \right)^{T} \mathcal{I} \left(C_i^6 - C_{i,min}^6\right),
\eea
where $C_{i,min}^6$ corresponds to the Wilson coefficients vector minimizing the $\chi^2$ and $ \mathcal{I} $ is the Fisher information matrix.
The resulting fit space of the $C_i^6$ is highly correlated\cite{Berthier:2015oma,Berthier:2015gja,Berthier:2016tkq}, with recent
results\cite{Brivio:2017bnu} shown in Fig.\ref{Fig:correlation_matrices}.
The effect of modifying the input parameters: $\{\hat{\alpha}_{ew},\Delta \hat{\alpha} \} \rightarrow \hat{m}_W$
was been recently examined \cite{Brivio:2017bnu}, which does not change this conclusion.
The Fisher matricies of the SMEFT fit space allow the constuction of the SMEFT $\chi^2$ function.
These matricies were developed in a fit using 177 observables\cite{Berthier:2015oma,Berthier:2015gja,Berthier:2016tkq,Brivio:2017bnu} and are available upon request.

\subsection{Characterizing and testing for measurement bias}\label{subsec:bias}
The results shown in Fig.\ref{Fig:correlation_matrices} were obtained in a global fit in the limit
$\Delta_{i,SMEFT} \rightarrow 0$, but theoretical errors exist in the SMEFT.
These theoretical errors have a number of sources:
\begin{itemize}
\item{The novel interactions present can bias the projection of a measurement onto the $C_i^6$ space.}
\item{The neglect of higher order terms in the SMEFT operator expansion introduces
a truncation error when combining data sets.}
\item{The scale dependence of the SMEFT operators, and the neglect of loop corrections
involving these operators, introduces a truncation error combining data sets.}
\end{itemize}
For point one, non-SM physics effects are limited to an analytic and local form by the Taylor expansion leading to the SMEFT, and can be examined
to characterize the leading sources of measurement bias.
The conclusion is that this bias is under control in EWPD pole measurements.\cite{Berthier:2015oma,Bjorn:2016zlr}
This illustrates the power of the SMEFT to develop model independent conclusions.
\begin{figure*}
\centering
\includegraphics[width=0.49\textwidth]{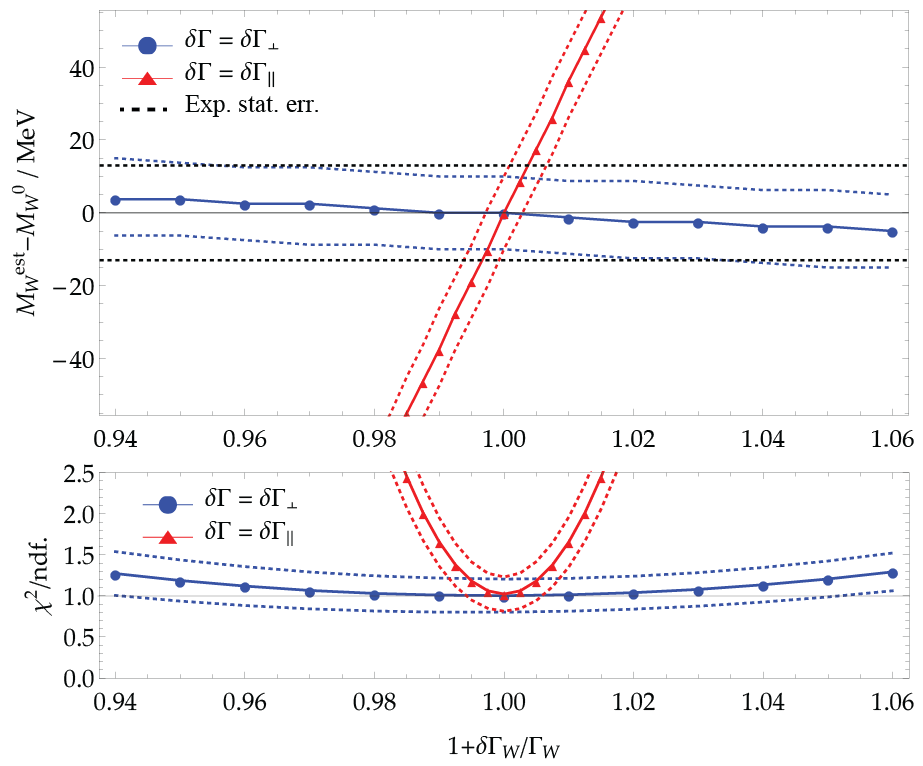}
\includegraphics[width=0.49\textwidth]{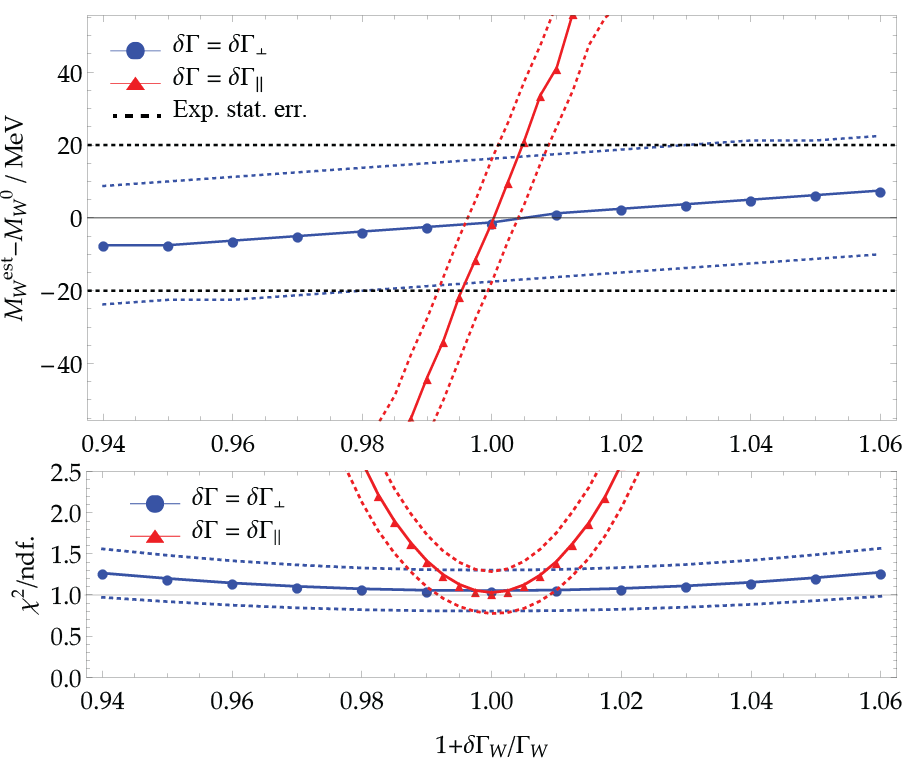}
\caption{The bias $m_W^{est}-m_W^0$ on the estimated $W$-mass relative to input mass in a fit to
(a) the $m_T$-distribution and (b) the $p_{T\ell}$-distribution, due to the presence of SMEFT operators.
The SMEFT contribution is decomposed into $\delta \Gamma_{||}$ and $\delta \Gamma_{\perp}$.
Note that this 1-d scan of the parameter space is only an approximation to a multi-dimensional parameter
scan varying $\hat{m}_W, \hat{\Gamma}_W$ simultaneously. \label{fig:results}}
\end{figure*}
In the case of LEP $Z$ pole data, the challenge is due to the projection of
LEP constraints onto the local contact operators appearing in tree level modifications of
$\bar{\psi} \psi \rightarrow Z \rightarrow \bar{\psi} \psi$, while neglecting the interference with
$\bar{\psi} \psi \bar{\psi} \psi$ operators.
If LEP $Z$ pole data was defined exactly on the $Z$ resonance peak,
this interference is known to vanish\cite{Berthier:2015oma,Han:2004az}. However, LEP
data combines $40 \, {\rm pb}^{-1}$ of off-peak data with $155 \, {\rm pb}^{-1}$ of on-peak data\cite{ALEPH:2005ab}. The interference effects due to $\bar{\psi} \psi \bar{\psi} \psi$ scale
as $\sim (m_Z \, \Gamma_Z/v^2)$ times a function of this ratio of off/on peak data\cite{Berthier:2015oma}.
The corresponding uncertainty does not disable using EWPD to obtain $\sim \%$ level constraints on the $C^6_i$.

In the case of $m_W$\cite{D0:2013jba,Aaltonen:2013vwa}, the SMEFT can be used to decompose the perturbations
due to local contact operators into directions perpendicular and parallel to the overall normalization of the
transverse variable spectra used to extract $\hat{m}_W$. The measurements are done choosing to have a floating normalization.
The SM theoretical errors \cite{Bjorn:2016zlr} dominate the measurement bias due to this choice,
as shown in Fig.\ref{fig:results}.
This extends a SM error analysis of these measurements\cite{CarloniCalame:2016ouw} to a model
independent conclusion\cite{Bjorn:2016zlr}.

\section{Loop corrections to $Z$ decay}\label{subsec:loops}
The SMEFT allows one to combine data sets into
a global constraint picture, but one loop calculations in this theory are required
to gain the full constraining power of precisely measured observables.
In the case of $\mathcal{O}(y_t^2,\lambda)$ corrections to $\{\Gamma_Z, \Gamma_{Z \rightarrow \bar{\psi \psi}},\Gamma_Z^{had},R_\ell^0,R_b^0\}$, about
thirty loops were determined\cite{Hartmann:2016pil} mapping the input parameters to these observables, while retaining the $m_t,m_h$ mass scales in the calculation.
The renormalization of the $\mathcal{L}_6$ operators in the warsaw basis\cite{Grojean:2013kd,Jenkins:2013zja,Jenkins:2013wua,Alonso:2013hga} is
used in this result, which simultaneously provides a check of the terms $\propto y_t^2,\lambda$ that appear in
these observables\cite{Hartmann:2016pil}. These calculations define a perturbative expansion of the observables used in EWPD
\bea
\bar{O}_i = \bar{O}^{LO}_i(C_i^6) + \frac{1}{16 \, \pi^2}\left(F_1[C_j^6]
+ F_2[\lambda,C_k^6]\log{\frac{\mu^2}{\hat{m}_h^2}}
+ F_3[y_t^2,C_l^6]\log{\frac{\mu^2}{\hat{m}_t^2}}\right) + \cdots
\eea
The LO results depend on ten Wilson coefficients in the warsaw basis, defining $C_i$, and ${\rm dim}(C_j)\neq {\rm dim}(C_k) \neq {\rm dim}(C_l) >{\rm dim}(C_i)$ in general.
At one loop, considering $\mathcal{O}(y_t^2,\lambda)$ corrections the following new SMEFT parameters appear\cite{Hartmann:2016pil}
\bea
\{C_{qq}^{(1)}, C_{qq}^{(3)}, C_{qu}^{(1)},C_{uu}, C_{qd}^{(1)},C_{ud}^{(1)},C_{\ell q}^{(1)},C_{\ell q}^{(3)},
C_{\ell u},C_{qe},C_{eu},C_{Hu}, C_{HB}+ C_{HW},C_{uB},C_{uW},C_{uH} \}.
\eea
The number of parameters exceeds the number of EWPD measurements.
EWPD is important to incorporate into the SMEFT as for a few observables $\Delta^{exp}_j \sim 0.1 \%$.
When
\bea
\frac{1}{16 \, \pi^2}\left(F_1[C_j^6]
+ F_2[\lambda,C_k^6]\log{\frac{\mu^2}{\hat{m}_h^2}}
+ F_3[y_t^2,C_l^6]\log{\frac{\mu^2}{\hat{m}_t^2}}\right) + \cdots \gtrsim \Delta^{exp}_j \hat{O}_i,
\eea
these corrections can have a significant effect on the interpretation of EWPD.
If this is the case depends on the values of the UV dependent Wilson coefficients.
For $Z$ pole EWPD measurements, one can fix $\mu = \hat{m}_Z$, but the new parameters are still present.
In principle, EFT techniques can sum all of the logs that appear relating various scales.
However, the extraction and prediction of EWPD in the SMEFT is a multi-scale problem
$ 0 \ll \hat{m}_\mu^2 \ll \hat{m}_Z^2 < \hat{m}_h^2 < \hat{m}_t^2$
and this requires fairly epic calculations be performed.
Mapping the Wilson coefficients to the matching scale $\mu \sim \Lambda$
one can infer the degree of constraint on the underlying theory.
Recent results are renormalized at the scale
$\mu \sim \Lambda$ to allow a direct examination of this question.\cite{Hartmann:2016pil}

A number of technical hurdles were overcome in this calculation.\cite{Hartmann:2016pil}
For example, evanescent scheme dependence resulting from
defining $\gamma_5$ in various ways in $d$ dimensions appeared in a novel manner.
A number of further technical hurdles remain in the way of determining the remaining one loop corrections to the full set of EWPD observables,
so our conclusions are limited to the partial $\mathcal{O}(y_t^2,\lambda)$ results known.
These results establish that LEP data does not constrain the SMEFT parameters appearing at tree level
in EWPD to the per-mille level in a model independent fashion.

When using the constraints resulting from EWPD to study LHC data, one must run the
determined constraints on $C_{i,j,k,l}(m_Z)$ to the various LHC measurement scales. This ``fuzzes'' out the constraints due to EWPD
when mapping between the data sets
by renormalization group equation (RGE) running. It is not advisable to set  $C_i^6(\mu) = 0$ in LHC analyses to attempt to incorporate EWPD constraints
for all of these reasons.
Doing so introduces inconsistencies which defeats the purpose of the SMEFT approach.
The challenge of combining EWPD with Higgs data requires further development of the SMEFT. The results discussed here are part of a one loop revolution
in SMEFT calculations\cite{Hartmann:2015oia,Hartmann:2015aia,Ghezzi:2015vva,Gauld:2016kuu,Gauld:2015lmb,Bylund:2016phk}.

\section{SMEFT reparameterization invariance}\label{subsec:reparam}
Even with the emergence of the SMEFT over the last few years,
the existence of unconstrained directions in certain operator bases
(when considering EWPD) has caused enormous confusion.  The physics of these unconstrained directions is
now understood in an operator basis independent manner.\cite{Brivio:2017bnu}
A massive vector boson can always be transformed between canonical and non-canonical form
in its kinetic term by a field redefinition without physical effect,
due to a corresponding correction in the LSZ formula. Such a shift can be canceled by a corresponding shift in the
$V \bar{\psi} \psi$ coupling. The same set of physical
scatterings can be parameterized by an equivalence class of fields and coupling parameters in the SMEFT
as a result\cite{Brivio:2017bnu}
 \bea\label{WOW}
 \left(V,g \right) \leftrightarrow \left(V' \, (1+ \epsilon), g' \, (1- \epsilon) \right),
 \eea
where  $\epsilon \sim \mathcal{O}(v^2/\Lambda^2)$.
We refer to this as SMEFT reparameterization invariance.
Denoting $\langle \cdots \rangle_{S_R}$ as the class of $\bar{\psi} \psi \rightarrow V \rightarrow  \bar{\psi} \psi$ matrix elements, the following operator
relations follow from the SM EOM in Eqn.\ref{EOMsm} (here $y_i$ denotes the hypercharge of state $i$)
\bea\label{EOMrelations}
\langle y_h \, g_1^2 Q_{HB} \rangle_{S_R} &=& \langle \sum_{\psi} y_k \, g_1^2 \, \overline \psi_\kappa \, \gamma_\beta  \psi_\kappa \, (H^\dagger \, i\overleftrightarrow D_\beta H)
+   2 \, g_1^2 \,Q_{HD}
- \frac{1}{2} g_1 \, g_2 \, Q_{HWB}  \rangle_{S_R}, \\
\langle  \, g_2^2 Q_{HW} \rangle_{S_R} &=& \langle  g_2^2 \,(\overline q \, \tau^I \gamma_\beta  q + \overline l \, \tau^I \gamma_\beta  l ) \, (H^\dagger \, i\overleftrightarrow D_\beta^I H)
- 2 \, g_1 \, g_2 \, y_h \, Q_{HWB}  \rangle_{S_R}.
\eea

Because of the reparameterization invariance, a Wilson coefficient multiplying the left hand side of these equations is
not observable in $\bar{\psi} \psi \rightarrow \bar{\psi} \psi$ scatterings.
The invariance of $S$ matrix elements under field configurations equivalent by use of the EOM  means
the corresponding fixed linear combinations of Wilson coefficients that appear on the right-hand sides of these equations
are also not observable in the $S_R$ matrix elements. The $S_R$ class of data is simultaneously invariant under the two independent reparameterizations (defining $w_{B,W}$)
that leave the products $(g_1 B_\mu)$ and $(g_2W^i_\mu)$ unchanged.
The unconstrained directions in the global fit, developed as described in Section \ref{subsec:generalities}
in the $\{\hat{\alpha}, \hat{m}_Z,\hat{G}_F\}$ input scheme, are found to be
\bea\label{empiracalflatalpha}
w_1= \frac{v^2}{\Lambda^2} \left(\frac{1}{3}C_{Hd}-2C_{HD}+ C_{He}+\frac{1}{2} \CHls-\frac{1}{6} \CHqs-\frac{2}{3} C_{Hu}-1.29 (\CHqt+ \CHlt)+ 1.64 C_{HWB}\right),\\
w_2=\frac{v^2}{\Lambda^2} \left(\frac{1}{3}C_{Hd}-2C_{HD}+ C_{He}+\frac{1}{2} \CHls-\frac{1}{6} \CHqs-\frac{2}{3} C_{Hu}+ 2.16 (\CHqt+ \CHlt)- 0.16 C_{HWB}\right).
\eea
These unconstrained directions have their origin in SMEFT reparameterization invariance, as they
can be projected into the vector space defined by $w_{B,W}$ as\cite{Brivio:2017bnu}
\bea\label{basisdecomp}
w_1 = -w_B - 2.59 \, w_W, \quad \quad w_2= -w_B +4.31 \, w_W.
\eea

\section{The non-minimal character of the SMEFT}\label{subsec:nonminimal}
LHC data is now enabling a SMEFT approach to physics beyond the SM.
Does it nevertheless make sense to only retain a few operators, not a general SMEFT, in a global analysis?
It is interesting to avoid fine tuned cases when examining the question of how extensively the SMEFT can be reduced to
a smaller subset of operators. Considering new physics sectors approximately
respecting the global symmetry group $\rm U(1)_B \otimes U(1)_L \otimes \rm SU(3)^5$,
and a discrete $\rm CP$ symmetry to accommodate flavour and EDM data,
a study of the non-minimal character of the SMEFT finds\cite{Jiang:2016czg}
\begin{itemize}
  \item{The RGE of the SMEFT\cite{Grojean:2013kd,Jenkins:2013zja,Jenkins:2013wua,Alonso:2013hga} indicates that the full theory
  should be used
  in a consistent analysis considering one loop effects.}
\item{To reduce the operator profile in the SMEFT in tree level matchings, heavy field content charged under the
$\rm SM$ gauge groups with non-trivial representations is generally required.}
\item{Heavy fermion fields generate a large number of operators matching to the SMEFT.
Heavy vector fields with nontrivial $\rm U(1)_Y$ charges, forbid the three point vector self interaction.
As a result, these vectors have a cut off scale proximate to their introduced mass as scattering amplitudes
of these vectors scale as $\mathcal{A} \propto s^2/m_V^4$, leading to the expectation
of a large number of operators due to non-perturbative matchings. Heavy scalars can generate the single operator $(H^\dagger H)^3$,
but if a mechanism is required to generate the heavy mass scales in the UV sector, multiple operators
also result.}
\end{itemize}

\subsection{Do universal theories exist?}
A universal theory assumption\cite{Barbieri:2004qk} is sometimes invoked as
an alternative to a SMEFT analysis.\footnote{The author is aware of Hinchliffe's rule.}
One can reexamine the idea of universal theories using the arguments developed examining the
non-minimal character of the SMEFT. The non-minimal character of the SMEFT\cite{Jiang:2016czg} largely follows from
demanding a mechanism be defined to generate UV masses $\Lambda \gg v$, so that a consistent IR limit
can be defined for matching.

A fully defined mechanism of dynamical mass generation in a UV sector has never been demonstrated to be
consistent with the universal theory assumption.\footnote{Nonuniversal effects are present in extra-dimensional scenarios\cite{Banks:1999eg}. These effects
depend on the fermion and Higgs embedding in the extra-dimensions
in a sensitive manner\cite{Banks:1999eg,Strumia:1999jm} allowing
a vanishing of such effects in a rather particular limit\cite{Strumia:1999jm}.
As a result, these corrections are sometimes dropped from EWPD analyses in these frameworks
and these theories are argued to be universal in character.}
As a specific example, universal theories have been motivated by considering the coupling
of $B_\mu',W_\mu'$ states to the full gauge currents on the right hand side of Eqn.4,5.
Retaining the operators $(\left[D^\alpha , W_{\alpha \beta} \right]^I)^2, (D^\alpha B_{\alpha \beta})^2$ in an $\mathcal{L}_6$
basis, then captures a tree level universal effect at $\mu \sim \Lambda$.
Non-perturbative matching effects due to a strongly interacting $B_\mu',W_\mu'$ mass generation sector,
including non-universal effects, can also be generated at this scale.
If a UV Higgs mechanism is invoked to generate the $B_\mu',W_\mu'$ masses, one can
 study a limit where this Higgs$'$ state is integrated out,
generating a UV chiral Lagrangian to embed the $B_\mu',W_\mu'$ states in, and subsequently match to the lower energy EFT.
Operators characterizing non-universal effects are present, and
the assumed embedding of the SM fermions in the UV sector plays a central role in determining
the scaling of the Wilson coefficients. A proof that
non-universal effects can be neglected in a well defined framework where the $B_\mu',W_\mu'$ masses
are dynamically generated is not available in the literature.

In any case, assumed
universal theories generate non-universal theories\cite{Trott:2014dma,Wells:2015cre} using the renormalization group to
run the operators matched onto from $\Lambda \rightarrow \hat{m}_z$.

\section{Conclusions}
The SMEFT is a theory of SM deviations that allows the study of LHC data in a unified framework with EWPD,
and other lower energy measurements.
This framework is systematically improvable and requires further development to consistently combine
EWPD and LHC data. This theory is undergoing a rapid development, some of which was reviewed here.

\section*{Acknowledgments}

I acknowledge L. Berthier, M. Bj\o rn, I. Brivio,
C. Hartmann, Y. Jiang and W. Shepherd for collaboration and
the Villum Fonden and the DNRF (DNRF91) for financial support.
Thanks to the organizers of Moriond EW 2017
for the opportunity to present these results.

\end{document}